\documentclass[
    ,final            
  ]
  {aipproc}

\layoutstyle{6x9}

\usepackage{graphicx}
\graphicspath{{converted_graphics/}}
\begin{document}

\title{Observations of Very Metal-Poor Stars\\ in the Galaxy}


\author{Timothy C. Beers}{
  address={Department of Physics and Astronomy, Center for the Study of
Cosmic Evolution, and Joint Institute for Nuclear Astrophysics, Michigan
State University, E. Lansing, MI  48824 USA}
}



\begin{abstract}
%


I report on recent results from observations of stars with metallicities [Fe/H]
$\le -2.0$. These include a substantial new sample of objects with
high-resolution observations obtained as part of a follow-up of the HK Survey,
The Hamburg/ESO Survey, and the ongoing survey SEGUE: Sloan Extension for
Galactic Understanding and Exploration.  Perspectives on the next directions are 
also provided.

\end{abstract}

\maketitle


\section{The Importance of Very Metal-Poor Stars}

It is widely recognized that stars with metallicities less than 1\% of the solar
value ([Fe/H] $ < -2.0$; the Very Metal-Poor (VMP) stars in the nomenclature of 
\citet{Beers2005} provide invaluable information on the origin of the
elements in the Universe, and on the nature of the neutron-capture processes
that were the dominant source of the elements beyond the iron peak. Perhaps less
appreciated is the fact that these same stars place strong constraints on the
formation and evolution of individual galaxies, and thereby provide a direct
connection with cosmological studies. Within the Milky Way, the shape of the
low-metallicity tail of the Metallicity Distribution Function (MDF) is beginning
to reveal the characteristic abundances of elements associated with the major
epochs of star formation in the early Galaxy. Changes in the MDF as a function of
distance are revealing the assembly history of the Galaxy (see, e.g., the
article by Carollo \& Beers in this volume). The frequency of various abundance
signatures, e.g., the [C/Fe] and [$\alpha$-element/Fe] ratios, are being used to
investigate the dominant production sites, and the Initial Mass Function (IMF)
of the stellar populations in the proto-Galactic fragments that theory suggests
were involved in galaxy assembly. In this short review, I summarize a few of the
recent observational results, and indicate the directions which the field will
be going in the next few years.

\section{Past and Ongoing Surveys}

Past searches for VMP stars include those based on the selection of high
proper-motion stars in the local volume \citep[e.g.,][]{Ryan1991,Carney1996},
and in particular, stars identified from non-kinematically biased {\it
in-situ} objective-prism surveys, such as the HK survey of Beers and colleagues
\citep{Beers1985,Beers1992}, and the more recent Hamburg/ESO Survey (HES) of
Christlieb and collaborators \citep{Christlieb2003}. These two surveys form the basis
for the majority of the high-resolution spectroscopic studies that have been
carried out over the past decade, such as the First Stars program of \citet{Cayrel2004},
the HERES project of \citet{Barklem2005}, and the 0Z Project of
\citet{Cohen2007}.

Although work continues on the medium-resolution follow-up of HK and HES
candidate VMP stars, to date the efforts have yielded roughly 2500 firm
identifications.  However, it should be emphasized that as these stars are
selected in the volume sorrounding the Sun, and explore no more than 10 kpc (in
the case of the HK survey) to 15 kpc (in the case of the HES) away, these
samples are dominated by the now-recognized inner-halo population of the Galaxy
\citep{Carollo2007}.  The outer-halo population, which Carollo et al. suggest
is comprised of a MDF with peak metallicity a factor of three times lower
([Fe/H]$ = -2.2$) than that associated with the inner halo ([Fe/H]$ = -1.6$), 
dominates outside of 15-20 kpc from the Galactic center. 
This may well account for the difficulty in the identification of significant
numbers of stars at the lowest metallicities, the Ultra Metal-Poor (UMP) and
Hyper Metal-Poor (HMP) stars, with [Fe/H] $< -4.0$ and [Fe/H] $< -5.0$,
respectively.  If progress is to be made in finding large numbers of
such stars, different tactics must be explored. 

Fortunately, this process has already begun.  Already, the largest numbers of
VMP stars revealed to date has emerged from medium-resolution spectroscopic
spectra taken during the course of the SDSS (primarily calibration stars) and
SEGUE.  SEGUE is a survey directed at studies of Galactic structure, but it
includes a number of target categories (such as F turnoff stars, K giants, and a
low-metallicity category extending across all spectral types) that is successful
in the identification of large numbers of VMP stars.  Even though SEGUE is not
yet complete, the total list of VMP stars already exceeds 10,000 stars,
quadruple the number from the HK/HES efforts combined. Of greatest importance,
SDSS/SEGUE can explore to much larger distances, beyond the 10-15 kpc region
where the inner-halo population dominates. Furthermore, the availability of
reasonably accurate proper motions for the subset of nearby stars (out of 4-5
kpc from the Sun) makes it feasible to identify candidate VMP, EMP, UMP, and HMP
stars based on the characteristic retrograde signature of the outer-halo
population. This mode of selection has not been implemented in SEGUE, but it is
one of the target categories that will be explored in the SEGUE-2 project, which
will be executed from July 2008 to July 2009 as part of the proposed next
extension of the Sloan Survey, known as SDSS-III (see \url{http:
//www.sdss3.org/outermilkyway.php}). We are hopeful that this dedicated effort
will finally be able to break through the UMP/HMP barrier.

Space precludes a comprehensive discussion of even the most recent results
coming from high-resolution spectroscopic studies of VMP stars, so below I
simply summarize a few of the results of greatest relevance to this conference.
Other examples can be gleaned from the online version of my talk.

\section{Recent Results}

\subsection{Carbon-Enhanced Metal-Poor (CEMP) stars}

The CEMP stars (and their sub-categories) are defined by \citet{Beers2005}
based on the criteria listed in Table 1. Their clear signficance to
understanding the chemical evolution of the Galaxy results from the apparently
large fraction of such stars that are found among the VMP stars (which varies
between 10\% and 20\%, according to investigations carried out to date). At the
lowest iron abundances, CEMP stars represent 40\% of stars with [Fe/H] $< -3.5$
\citep{Beers2005}, and 100\% (3 of 3) of stars known with [Fe/H] $<
-4.0$ \citep{Christlieb2002,Frebel2005,Norris2007}. A number
of articles in this volume discuss attempts to understand the origin of the UMP
and HMP CEMP stars.   

\bigskip
\bigskip
\begin{table}[h]
\begin{tabular}{ll}\hline
  \multicolumn{2}{l}{Carbon-enhanced metal-poor stars}\rule{0ex}{2.3ex} \\\hline
  CEMP      & $\mbox{[C/Fe]} > +1.0$\rule{0ex}{2.3ex}\\
  CEMP-r    & $\mbox{[C/Fe]} > +1.0$ and $\mbox{[Eu/Fe]} > +1.0$\\
  CEMP-s    & $\mbox{[C/Fe]} > +1.0$, $\mbox{[Ba/Fe]} > +1.0$, and $\mbox{[Ba/Eu]} > +0.5$\\
  CEMP-r/s  & $\mbox{[C/Fe]} > +1.0$ and $0.0 < \mbox{[Ba/Eu]} < +0.5$\\
  CEMP-no   & $\mbox{[C/Fe]} > +1.0$ and $\mbox{[Ba/Fe]} < 0$\\\hline
  \multicolumn{2}{l}{Neutron-capture-rich stars}\rule{0ex}{2.3ex} \\\hline
  r-I   & $+0.5 \le \mbox{[Eu/Fe]} \le +1.0$ and $\mbox{[Ba/Eu]} < 0$\rule{0ex}{2.3ex}\\
  r-II  & $\mbox{[Eu/Fe]} > +1.0$ and $\mbox{[Ba/Eu]} < 0$\\
  s     & $\mbox{[Ba/Fe]} > +1.0$ and $\mbox{[Ba/Eu]} > +0.5$\\
  r/s   & $0.0 < \mbox{[Ba/Eu]} < +0.5$  \\\hline
\end{tabular}
\caption{Definition of sub-classes of metal-poor stars (based on \citet{Beers2005}
}
\end{table}
\bigskip

\citet{Aoki2007} reports new observations (obtained with the Subaru telescope)
and summarizes results from the high-resolution studies of CEMP stars reported
in the literature. Among the most intriguing results from this study concerns
the apparent difference in the MDFs (see Figure 1) between the CEMP-s (those
exhibiting abundance signatures characteristic of the s-process, and which
likely originated from mass-tranfer events from a now-deceased intermediate mass
AGB companion) stars, and the CEMP-no (those exhibiting no neutron-capture
elements, and whose origin remains the subject of current discussion) stars.

\begin{figure}[tbp] 
  \centering
  \includegraphics[width=2.84in, height=2.55in]{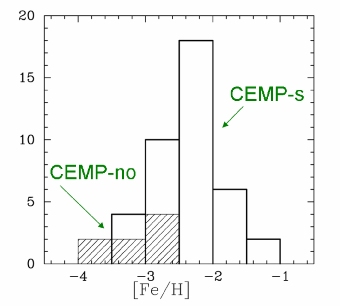}
  \caption{Distribution of [C/Fe] for CEMP stars discussed by \citet{Aoki2007}.  Note
   the predominance of the CEMP-no stars (cross-hatched bars) at the lowest [Fe/H].}
  \label{fig:beers_figure1}
\end{figure}

\citet{Aoki2008} presents additional Subaru results for a small sample
(seven) of CEMP stars selected from SDSS/SEGUE to be located at the
main-sequence halo turnoff.  These stars are important because they lie in the
region of the H-R diagram where there is no possibility of extensive mixing of
their outer atmospheres having taken place (at least via usual mechanisms); they
thus represent an essentially pure signature of the elemental abundance patterns
produced by their progenitors.  It is of interest that the majority of the CEMP
stars in this study exhibit kinematics that may associate them with the
outer-halo population reported by \citet{Carollo2007}.

Another recent study has measure the abundance of flourine (F) in a CEMP-s star
\citep[see Figure 2]{Schuler2007}, using the Phoenix near-IR high-resolution
spectrograph on the Gemini-S telescope. This observation represents the first
such detection in a CEMP star (and revealed this star to have an over-abundance
[F/Fe] $= +2.9$, nearly 1000 times that of the solar ratio).  This element
provides a senstive probe of the operation of the s-process in AGB stars
\citep[e.g.,][]{Campbell2007}, and may provide information on the IMF from which the AGB
progenitor was drawn \citep{Lugaro2008}.  Additional high-resolution
spectroscopy of CEMP stars for the study of this element are clearly required.

\begin{figure}[h] 
  \centering
  \includegraphics[width=2.27in,height=2.31in]{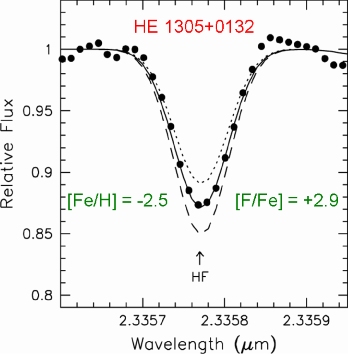}
  \caption{The detection of F in a VMP CEMP star by \citet{Schuler2007}; the HF molecular feature is very prominent.}
  \label{fig:beers_figure2}
\end{figure}

Among the recent theoretical studies of the significance of the CEMP stars, I
draw attention to the work of \citet{Tumlinson2007}, which has argued that the
critical mass of early star formation in the Universe was shifted to favor the
formation of intermediate-mass stars (essentially due to the ``floor'' set by
the cosmic microwave background temperature at the time) that are copius carbon
producers. Others \citep[e.g.,][]{Meynet2006,Chiappini2007}
have argued that significant carbon (and nitrogen) may be produced from the
winds of massive Mega Metal-Poor (MMP; [Fe/H] $< -6$), rapidly rotating stars
that might have formed at the very earliest epochs of the Universe. Such stars
would not be expected to produce s-process elements, but their association with
the origin of the CEMP-no stars has yet to be established.
 
\subsection{Highly r-process-enhanced stars}

One of the goals of the HERES survey of \citet{Barklem2005} was to
dramatically increase the numbers of VMP stars with enhanced r-process elements.
This goal was met; eight new examples of the phenomenon were identified, and a
critical difference was noted between the MDF of the so-called r-II stars (those
with [Eu/Fe] $> +1.0$, and low Ba, according to Table 1), and
the moderately r-process-enhanced stars ($+0.5 \le$ [Eu/Fe] $< +1.0$). As is clear
from Figure 3, while the r-I stars exhibit an MDF that covers a wide range, the
MDF of the r-II stars is restricted to roughly $-3.2 <$ [Fe/H] $< -2.6$).

\begin{figure}[tbp] 
  \centering
  \includegraphics[width=3.69in,height=3.07in]{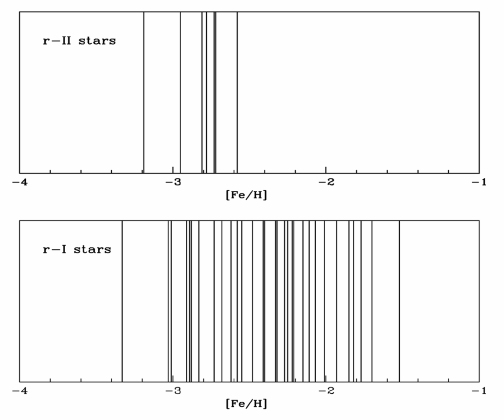}
  \caption{Distribution of [Fe/H] for r-process-enhanced stars identified by \citet{Barklem2005}.  Note
   the obvious contrast between the behavior of the r-II stars with that of the r-I stars.}
  \label{fig:beers_figure3}
\end{figure}

An additional handful of r-II stars have since been identified by other studies,
including at least one with detected uranium (\citep{Frebel2007}; HE~1523-0901,
with [Fe/H] $=-2.95$); others are still unpublished. It is presumably no
coincidence that {\it all} of the newly discovered r-II stars fall in the same
low metallicity interval as those from the Barklem et al. study. One possible
explanation is that the main astrophysical r-process arises from from stars of a
relatively small mass range (many have suggested 10-15 M$_\odot$), which may
have been the dominant contributor of heavy elements at the time when the Galaxy
reached a mean metallicity around [Fe/H] $\sim -3.0$. Other alternatives surely
exist.

\subsection{Observations of light n-capture elements}

Although the stars with large enhancements of n-capture elements are of great
interest, it is equally important to consider the abundances of n-capture
elements for the great majority of metal-poor stars with ``normal'' levels, as
these help constrain the question of the universality of the r-process, and
address whether multiple r-process sites must be considered.  Recently, \citet{Francois2007}
reported abundance determinations for some 16 n-capture elements
for a sample of 32 VMP and EMP stars observed during the course of the Cayrel et
al. First Stars program.  These data greatly enlarge the numbers of stars with
[Fe/H] $< -2.8$ with such measurements available.  The star-to-star scatter
among these elements reaches a peak at around [Fe/H] $ = -3.0$, but below this
metallicity, too few stars exist with measurements to be certain of the
behavior.

\begin{figure}[h] 
  \centering
  \includegraphics[width=2.84in,height=3.57in]{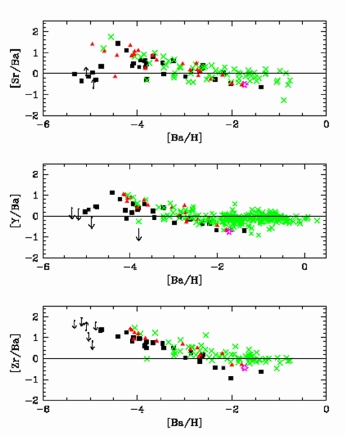}
  \caption{Reported abundances for Sr, Y, and Zr, relative to Ba, as a function of [Ba/H], 
from \citet{Francois2007}.  The new data are shown as the black squares (or upper limits).
Other data are from the literature.}
  \label{fig:beers_figure4}
\end{figure}

By adopting the element Ba as a reference element, Francois et al. demonstrate
the existence of strong anti-correlations of the lighter n-capture ratios
[Sr/Ba], [Y/Ba], and [Zr/Ba] with the [Ba/H] abundance from [Ba/H] $\sim -1.5$
down to [Ba/H] $\sim -4.5$ (Figure 4).  It remains unclear if this behavior
changes below [Ba/H] $ = -4.5$, as only upper limits on the lighter element
ratios are available for the majority of stars in this region; the few
measurements that {\it do} exist suggest a possible reversal of the trends
observed above this limit.  In any case, the results confirm that there must
exist a second n-capture process to account for the synthesis of the first
r-process peak elements, variously referred to in the literature as the ``weak''
r-process \citep{Qian2000} or the Light Element Production Process 
\citep[LEPP][]{Travaglio2004}.  Subtraction of the predicted contributions to
these elements by the main r-process suggests that this mechanism is responsible
for the production of 90-95\% of the observed amount of Sr, Y, and Zr in stars
with [Ba/H] $> -4.5$.  See \citet{Montes2007} for additional discussion of the
LEPP.

\section{New and Anticipated High-Resolution Spectroscopic Studies}

There exist a number of large high-resolution spectroscopic surveys that are
just now getting underway, which should greatly enlarge the numbers of VMP stars
with available elemental abundance information.  The Chemical Abundances of Halo
Stars (CASH) survey is making use of the Hobby-Eberly
Telescope to obtain moderately high-resolution ($R = 15,000$) spectroscopy for up to 1000 VMP
stars identified during the course of the HK-I, HK-II, and HES efforts, with a
large number of additional stars from SDSS/SEGUE.   Aoki et al. have recently
been awarded a Key Project status on the Subaru Telescope, with the aim of
obtaining $R=50,000$ spectroscopic observations for up to 200 VMP stars, the
majority of which will be drawn from SDSS/SEGUE targets.  Both of these surveys have as
one of their primary aims to test for the existence (or not) of chemical
signatures that might be associated with the inner/outer halo dichotomy reported
by \citet{Carollo2007}.  

In the near future, the proposed SDSS-III project will undertake a
massive survey of some 100,000 red giants with resolving power $R = 20,000$,
concentrating on the H-band region in the near-IR. This survey, known as APOGEE
(APO Galactic Evolution Experiment; see \url{http:
//www.sdss3.org/innermilkyway.php}) is expected to take place between 2011 and
2014, and will target objects in the Galactic bulge, bar, disk, and halo
components. It is expected that on the order of 15 individual elements will be
obtained per star.

Of course, we all look forward to the possible execution of the WFMOS (Wide
Field Multi-Object Spectrograph) survey of on the order of one million stars
at resolving power $R = 50,000$.  Currently, this Gemini instrument is expected
to be mounted on the prime focus of the Subaru telescope, in order to take
advantage of its wide field of view.  The hope, and expectation, is that this
survey will finally reveal the elemental abundances of stars that probe the
entire history of chemical evolution throughout the Galaxy.



\begin{theacknowledgments}

The author is grateful to the organizers for providing assistance with travel
and accomodation expenses.  This work also received support from grants AST
07-07776 and PHY 02-15783; Physics Frontier Center / Joint Institute for
Nuclear Astrophysics (JINA), awarded by the US National Science Foundation.

\end{theacknowledgments}

\bibliographystyle{aipprocl} 

\begin{thebibliography}{9}

\bibitem[Beers \& Christlieb (2005)]{Beers2005} T.~C. Beers \& N.~Christlieb, ``The Discovery and Analysis of Very
Metal-Poor Stars in the Galaxy,'' in \emph{Annual Reviews of Astronomy \&
Astrophysics}, \textbf{43}, 2005, pp. 531--580.

\bibitem[Ryan and Norris (1991)]{Ryan1991} S.~G. Ryan \& J.~E. Norris, \emph{Astron. J.} \textbf{101},
1835--1864 (1991).

\bibitem[Carney et al. (1996)]{Carney1996} B.~W. Carney et al., \emph{Astron. J.} \textbf{112},
668--692 (1996).

\bibitem[Beers et al. (1985)]{Beers1985} T.~C. Beers, G.~W. Preston, \& S.~A. Shectman,  \emph{Astron. J.}
\textbf{90}, 2089--2102 (1985).

\bibitem[Beers et al. (1992)]{Beers1992} T.~C. Beers, G.~W. Preston, \& S.~A. Shectman  \emph{Astron. J.}
\textbf{103}, 1987--2034 (1992).

\bibitem[Christlieb (2003)]{Christlieb2003} N. Christlieb, ``Finding the Most Metal-Poor Stars of the Galactic
Halo with the Hamburg/ESO Objective-Prism Survey,'' in \emph{Reviews of Modern
Astronomy}, \textbf{16}, 2003, pp. 191--206.

\bibitem[Cayrel et al. (2004)]{Cayrel2004} R. Cayrel et al., \emph{Astron. \& Astrophys.} \textbf{416},
1117--1138 (2004).

\bibitem[Barklem et al. (2005)]{Barklem2005} P.~S. Barklem et al., \emph{Astron. \& Astrophys.} \textbf{439},
129--151 (2005).

\bibitem[Cohen et al. (2007)]{Cohen2007} J.~G. Cohen et al., arXiv: 0709.1279 (2007).

\bibitem[Carollo et al. (2007)]{Carollo2007} D. Carollo et al., \emph{Nature} \textbf{450}, 1020--1025 (2007).

\bibitem[Christlieb et al. (2002)]{Christlieb2002} N. Christlieb et al., \emph{Nature} \textbf{419}, 904--906 (2002). 

\bibitem[Frebel et al. (2005)]{Frebel2005} A. Frebel et al., \emph{Nature} \textbf{434}, 871--873 (2005). 

\bibitem[Norris et al. (2007)]{Norris2007} J.~E. Norris et al., \emph{Astrophys. J.} \textbf{670}, 774--788 (2007).

\bibitem[Aoki et al. (2007)]{Aoki2007} W. Aoki et al., \emph{Astrophys. J.} \textbf{655}, 492--521 (2007).

\bibitem[Aoki et al. (2008)]{Aoki2008} W. Aoki et al., \emph{Astrophys. J.}, in press (2008)

\bibitem[Schuler et al. (2007)]{Schuler2007} S.~C. Schuler et al., \emph{Astrophys. J.} \textbf{667}, L81--L84 (2007).

\bibitem[Campbell \& Lattanzio (2007)]{Campbell2007} S.~W. Campbell \& J.~C. Lattanzio, arXiv: 0709:4567 (2007).

\bibitem[Lugaro et al. (2008)]{Lugaro2008} M. Lugaro et al., \emph{Astrophys. J.}, submitted (2008).

\bibitem[Tumlinson (2007)]{Tumlinson2007} J. Tumlinson, \emph{Astrophys. J.} \textbf{664}, L63--L66 (2007).

\bibitem[Meynet et al. (2006)]{Meynet2006} G. Meynet, S. Ekstrom, \& A. Maeder, \emph{Astron. \& Astrophys.}
\textbf{447}, 623--639 (2006) 

\bibitem[Chiappini et al. (2007)]{Chiappini2007} C. Chiappini et al., arXiv: 0712.3434 (2007). 

\bibitem[Frebel et al. (2007)]{Frebel2007} A. Frebel et al., \emph{Astrophys. J.} \textbf{660}, L117--L120 (2007).

\bibitem[Francois et al. (2007)]{Francois2007} P. Francois et al.  \emph{Astron. \& Astrophys.} \textbf{476}, 935--950 (2007). 

\bibitem[Qian \& Wasserburg (2000)]{Qian2000} Y.-Z. Qian \& G.~J. Wasserburg, \emph{Phys. Rep.} \textbf{333}, 77--108 (2000).

\bibitem[Travaglio et al. (2004)]{Travaglio2004} C. Travaglio et al., \emph{Astrophys. J.} \textbf{601}, 864--884 (2004).

\bibitem[Montes et al. (2007)]{Montes2007} F. Montes et al. 2007, \emph{Astrophs. J.} \textbf{671}, 1685--1695 (2007).

\end{thebibliography}

\end{document}